\documentclass[smallextended]{svjour3}       
\usepackage{graphicx}
\usepackage{hyperref}
\hypersetup{colorlinks=true,breaklinks,linkcolor=red,citecolor=blue}
\begin{document}

\title{Pairing and short-range correlations in nuclear systems}

\author{A. Rios \and
        A. Polls \and
        W. H. Dickhoff
}

\institute{A. Rios \at
              Department of Physics, Faculty of Engineering and Physical Sciences, University of Surrey, Guildford, Surrey GU2 7XH, United Kingdom \\
              \email{a.rios@surrey.ac.uk}
           \and
           A. Polls \at
              Departament de Física Qu{\`a}ntica i Astrof{\'i}sica and Institut de Ci{\`e}nces del Cosmos, Universitat de Barcelona, Avinguda Diagonal 647, E-8028 Barcelona, Spain\\ 
              \and
              W. H. Dickhoff \at
              Department of Physics, Washington University, St. Louis, Missouri 63130, USA
}

\date{Received: date / Accepted: date}

\maketitle

\begin{abstract}
The structure and density dependence of the pairing gap in infinite matter is relevant for astrophysical phenomena and provides a starting point for the discussion of pairing properties in nuclear structure. Short-range correlations can significantly deplete the available single-particle strength around the Fermi surface and thus provide a reduction mechanism of the pairing gap. Here, we study this effect in the singlet and triplet channels of both neutron matter and symmetric nuclear matter. Our calculations use phase-shift equivalent interactions and chiral two-body and three-body interactions as a starting point. We find an unambiguous reduction of the gap in all channels with very small dependence on the NN force in the singlet neutron matter and the triplet nuclear matter channel. In the latter channel, SRC alone provide a $50 \, \%$ reduction of the pairing gap.

\end{abstract}

\section{Introduction}
\label{intro}

Superfluidity plays an important role in nuclear physics. Nuclear structure is greatly influenced by pairing  \cite{Dean2003}. Nuclear reactions can be sensitive to pairing correlations \cite{broglia2013}. In nuclear physics, the Bardeen-Cooper-Schrieffer (BCS) approach, coupled to density functional techniques for nuclear structure, has been particularly successful in addressing a number of observed patterns that are sensitive to pairing correlations \cite{RingSchuck1980,Margueron2008}. There are still open questions in terms of how these patterns arise from first principles \cite{Soma2011,Soma2014}. Among others, we focus here on three key issues that affect the pairing gap in infinite nuclear matter. The first one is the role that strong short-range correlations (SRC) play in dampening the pairing gap. Second, what is the influence of three-nucleon forces (3NF) on pairing properties? Third, and final, one would like to know the size and density dependence of isoscalar neutron-proton pairing correlations. In this contribution, we will focus on these aspects and address them using infinite nuclear matter calculations based on a many-body approach that consistently treats SRC, 3NFs and tensor correlations. Long-range correlations can and will significantly affect the pairing gap, but we do not address them directly here \cite{dickhoff1987self,Shen2003,Shen2005,Ding2016}.

Short-range correlations have an impact on nuclear structure \cite{Wiringa2014,Atti2015,Neff2015,Dickhoff2004a}. A key effect of SRC is the removal of single-particle strength below the Fermi surface, which automatically brings in a population of high-momentum components \cite{Sargsian2014,Rios2009,Rios2014}. These components have been unambiguously identified in a variety of electron knockout reactions \cite{Rohe2004,Hen2014,Arrington2011}. The question of how this removal of strength affects the Fermi surface and its surroundings is technically difficult to address, but is capital for pairing properties which are sensitive to correlation effects \cite{Bozek1999,Baldo2000}. 
A variety of many-body techniques, like Brueckner--Hartree--Fock \cite{Shen2003} or Correlated Basis Functions methods \cite{Benhar2017}, have been extended to treat pairing correlations.
Here, we present results based on the self-consistent Green's function (SCGF) approach developed in Ref.~\cite{Muther2005} and further extended in Ref.~\cite{Ding2016}. The approach provides a fully microscopic account of single-particle strength removal based on realistic nucleon-nucleon (NN) interactions and, in its extended pairing formulation, is well suited to account for SRC in pairing properties.

Pairing properties can be directly linked to the underlying NN interaction \cite{Dean2003}. NN forces based on chiral perturbation theory have become a useful tool in nuclear ab initio studies \cite{Epelbaum2009,Machleidt2011}. A key advantage from this approach is that NN forces go hand in hand with associated 3NF. Besides that, chiral interactions also provide a way to explore uncertainties in many-body calculations, either by using cut-off variation \cite{Bogner2010} or more sophisticated techniques \cite{Epelbaum2015}. These approaches have recently been used to estimate systematic errors in pairing gaps within the BCS approach, including the effect of 3NFs~\cite{Srinivas2016,Drischler2017}. The SCGF method can be reformulated starting from NN and 3NF forces \cite{Carbone2013a}, providing a transparent procedure to incorporate 3NFs into many-body calculations considering SRC effects. In the case of chiral interactions, 3NFs at N2LO have been implemented \cite{Carbone2013b,Carbone2014} and provide results that agree well with the uncorrelated averages based on a free space normal ordering procedure \cite{Holt2010,Hebeler2010a,Srinivas2016,Drischler2017}. These interactions can be further incorporated into a pairing scheme and provide insight on the relevance of 3NF for pairing physics. 

Experimental and theoretical evidence points towards the fact that nuclear pairing phenomena can be explained using (neutron-neutron) $nn$ or (proton-proton) $pp$ pairing alone \cite{Dean2003}. In stark contrast, BCS predictions in infinite matter using realistic NN forces predict the largest pairing gap in the $^3$SD$_1$ channel \cite{Vonderfecht1991,Baldo1992,Baldo1995,Maurizio2014}. Isospin symmetric nuclei do not show any phenomenological signal of dominant, strong $np$ pairing \cite{Afanasjev2015}. This seems to point towards the fact that BCS predictions are invalid in this channel, at the very least for finite systems if not in infinite matter. Whatever quenching mechanism operates in this channel, then, it must be significant enough to strongly reduce attractive pairing correlations. We study the effect of the SRC depletion of the nuclear gap in symmetric matter using SCGF with 3NF.

The paper is organised as follows. Section~\ref{sec:formal} covers the formalism and theoretical approach. Results for neutron matter are given in Section~\ref{sec:neutron}, whereas symmetric matter and  $^3$SD$_1$ pairing are discussed in  Section~\ref{sec:symmetric}. Conclusions and an outline of future research are provided in Section~\ref{sec:conclusions}.

\section{Formalism}
\label{sec:formal}

In general, the BCS nuclear problem can be recast into an integral gap equation with the following form:
\begin{equation}
\Delta_k^{J(L)} = - \sum_{L'} \int \frac{\textrm{d}  k'}{(2 \pi)^3} 
\frac{i^{L-L'} V_{LL'}^{J}(k,k')}{2 \xi_{k'} } \Delta_{k'}^{J(L')} \, ,
\label{eq:gap}
\end{equation}
where $\Delta_k^J$ is the pairing gap for a given total pair angular momentum $\vec J=\vec L+\vec S$. In a BCS picture, $V_{LL'}^{J}(k,k')$ is the bare NN interaction and $k$ ($k'$) represents an incoming (outgoing) relative momentum for the usual BCS kinematics where center-of-mass momentum is set to zero. We consider coupled partial waves where $J$ is conserved but $L \neq L'$. The energy denominator $\xi_k$ contains the information of the in-medium propagation of paired particles. In the lowest order BCS approximation which we have implemented here
\begin{equation}
\xi_k = \sqrt{ (\varepsilon_k - \mu)^2 + \bar{\Delta_k}^2 } \, ,
\label{eq:den}
\end{equation}
where $\mu$ is the chemical potential. 
The quasi-particle spectrum $\varepsilon_k$ provides a dispersion relation between single-particle momenta and energies.
The gap equation above is derived under an angle average approximation, so that the average gap in the denominator is insensitive to the $L$ quantum number, $\bar \Delta_k^2= \sum_{L} \left[ \Delta_k^{J(L)} \right]^2$. 

 Within the Gorkov formalism \cite{Bozek1999}, the energy denominator of the gap equation is provided by a double-energy convolution, 
\begin{equation}
\frac{1}{2 \xi_k} = \int  \frac{\textrm{d}  \omega}{2 \pi}  \frac{\textrm{d}  \omega'}{2 \pi}  \frac{1-f(\omega)-f(\omega')}{\omega+\omega'}  A_k (\omega) A^s_k (\omega') \, .
\label{eq:fragden}
\end{equation}
Here, $f(\omega) = \left[ 1 + \exp \left( \frac{\omega-\mu}{T} \right) \right]^{-1}$ is a Fermi-Dirac distribution. The single-particle spectral function $A_k(\omega)$ describes the fragmentation of strength for a given momentum $k$ in the normal state \cite{Dickhoff08}. In a quasi-particle picture, it reduces to a single $\delta$ peak centered at a dressed quasi-particle energy, $\varepsilon_k$. The superfluid spectral function $A^s_k(\omega)$, in contrast, accounts for the presence of superfluidity. In a quasi-particle approximation, it describes a gapped spectrum and thus carries effectively information on the gap \cite{Bozek1999,Muther2005}. SRC associated with realistic NN forces induce fragmentation on the normal and superfluid spectral functions. The coupled self-consistent pairing problem including off-shell, energy-dependent spectral functions has only been solved with simplified separable interactions \cite{Bozek1999}.

Medium modifications to the BCS gap equation have been suggested to approximately account for the fragmentation effects associated with Eq.~(\ref{eq:fragden}). These can be broadly divided into two subsets. In the lowest order BCS approach, the single-particle energies in Eq.~(\ref{eq:den}) are given by kinetic energies, $\varepsilon_k=\frac{k^2}{2m}$. The first type of correction attempts to modify the single-particle energies by introducing a momentum-dependent single-particle in-medium shift, $U_k$. The overall effect can be very well approximated in terms of effective masses at the Fermi surface, $m^*$, and the gap is approximately quenched by a factor $m^*/m$ \cite{Baldo1995}. The second kind of correction attempts to take into account not only the shift, but also the reduction of single-particle strength around the Fermi surface. To avoid the double convolution, fragmentation is introduced via a momentum-dependent $Z$-factor, $Z_k < 1$ \cite{Shen2005,Bozek2003}. 
In the following, we will present results where these two effects are taken into account, and refer to it as the ``BCS+$\varepsilon_k$+$Z_k$" approximation.
Unfortunately, this approach only reproduces qualitatively the fragmentation of single-particle strength and in-medium quasi-particle shifts.

The fragmentation induced by SRC in infinite matter can be accessed directly by means of finite temperature SCGF calculations \cite{Dickhoff08}. In this approach, in-medium off-shell effects are consistently incorporated in the self-energy and in the effective T-matrix interaction \cite{Frick2003,Soma2008}. The single-particle spectral functions thus obtained have relatively broad, momentum-dependent quasi-particle energies as well as extended high-energy tails \cite{Rios2014}. Normal single-particle spectral functions can now be routinely obtained in infinite matter at finite density and temperature for any realistic NN force \cite{Ding2016,Rios2014,Carbone2014}. An extrapolation procedure that takes into account the thermodynamical consistency of the theory has been implemented to obtain the zero temperature normal spectral functions \cite{Ding2016}, which can then be convoluted to get a ``double normal" denominator
\begin{equation}
\frac{1}{2 \chi_k} = \int  \frac{\textrm{d}  \omega}{2 \pi}  \frac{\textrm{d}  \omega'}{2 \pi}  \frac{1-f(\omega)-f(\omega')}{\omega+\omega'}  A_k (\omega) A_k (\omega') \, .
\label{eq:fragden_normal}
\end{equation}
While this denominator does not contain any information on the superfluid, it does consistently account for the fragmentation of the spectral strength in infinite matter. It can therefore be used to inform the superfluid phase of this fragmentation. The first step towards a full self-consistent Gorkov solution is in fact found by including this double normal denominator in the gap equation: 
\begin{equation}
\Delta_k^{J(L)} = - \sum_{L'} \int \frac{\textrm{d}  k'}{(2 \pi)^3}
\frac{i^{L-L'} V_{LL'}^{J}(k,k')}{2  \sqrt{ \chi^2_{k'} + \bar{\Delta_{k'}}^2} } \Delta_{k'}^{J(L')} \, .
\label{eq:srcbcs}
\end{equation}
The gap contribution in the denominator guarantees the nonlinearity that is necessary to obtain a nontrivial pairing gap. The results obtained within this scheme differ qualitatively from those obtained in extended BCS approaches with effective single-particle energies, $U_k$, and/or strengths, $Z_k$ \cite{Muther2005,Ding2016}. In the following, we will present results obtained within this approach and label them ``SRC" to indicate that the associated gap calculations explicitly incorporate the effect of SRC.

Three-body forces play an important role in the saturation properties of infinite nuclear matter and arise naturally in chiral perturbation theory \cite{Epelbaum2009,Machleidt2011}. These can be incorporated into SCGF infinite matter calculations following the techniques outlined in Ref.~\cite{Carbone2013a} and implemented in infinite matter in Refs.~\cite{Carbone2013b,Carbone2014}. These represent an extension of the normal-ordering approach, but incorporating explicitly the effect of SRC via correlated momentum distributions. The effect of 3NF on single-particle properties below saturation is rather mild, whereas above saturation the quasi-particle spectrum becomes generally more repulsive. Depending on the 3NF, the momentum dependence can also change and hence induce  a substantial difference in effective masses. In the following, all SCGF calculations including 3NF have been obtained within the correlated approach with internal regulators~\cite{Carbone2014}. 

3NF can and have been incorporated in pairing calculations in the past \cite{Zuo2008,Dong2013}. Formally, one can prove from the original BCS formalism that 3NF enter the gap equation, Eq.~(\ref{eq:gap}), via an average of the third particle in the normal-ordering spirit \cite{papa2016}.  When fully antisymmetrized matrix elements are used in both the density-averaged 3NF, $W_{LL'}^{J}$, and the 2NF, $V_{LL'}^{J}$,  the two contributions are added together to obtain the bare pairing interaction in Eq.~(\ref{eq:gap}) \cite{Hebeler2010a,Srinivas2016,Drischler2017}. The pairing results we show have been obtained by solving the gap equation with these extended pairing interactions. The 3NF in this case is however reduced to a density-dependent force using an uncorrelated Fermi distribution. We do not expect that this small inconsistency affects the results in any qualitative way. The extension of the pairing approach to incorporate correlated averages is underway. 

\section{Pairing in neutron matter}
\label{sec:neutron}

We start our discussion by considering the case of pure neutron matter, in which isoscalar partial waves are forbidden. We present results for  the $^1$S$_0$ singlet channel, which dominates at low densities, first. We subsequently discuss the coupled triplet channel, $^3$PF$_2$, which is active at higher densities. Both cases are of relevance for astrophysics, particularly neutrino cooling \cite{Page2011,Ho2015}, although we refrain here from an at-length discussion as we do not include LRC effects. 

\subsection{Singlet pairing}
\label{sec:nsinglet}

\begin{figure}
  \includegraphics[width=\linewidth]{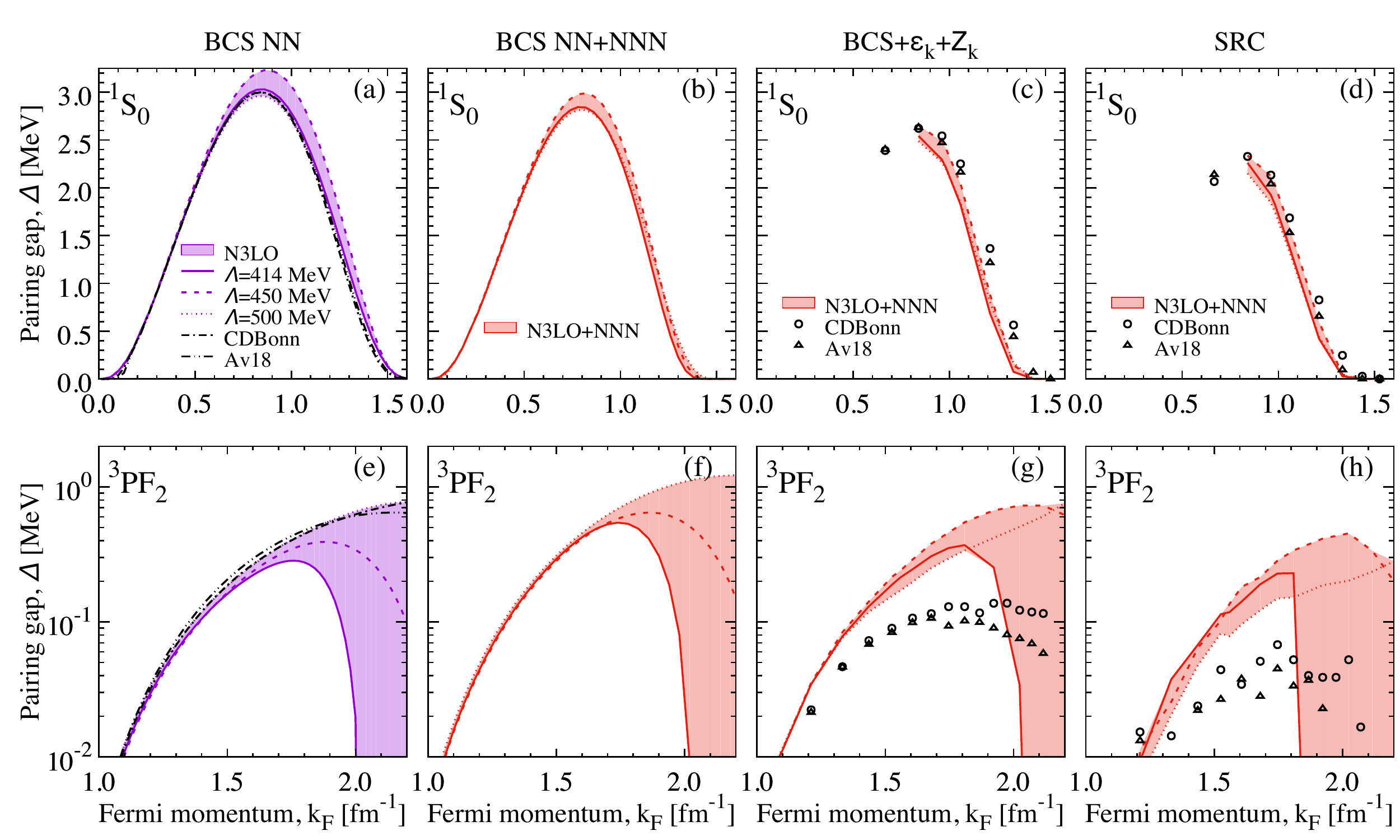}
\caption{Top panels:
singlet gap in neutron matter at the Fermi surface as a function of Fermi momentum, $k_F$. From left to right, panels correspond to (a) BCS results with NN forces only, (b) BCS results with chiral NN+3NF forces, (c) BCS with single-particle spectra with chiral NN+3NF forces and strengths and (d) SRC results with chiral NN+3NF forces, respectively. Bottom panels: the same for the triplet channel in neutron matter. }
\label{fig:gap_neumat}      
\end{figure}

We present the results of singlet pairing in neutron matter in the top panels (a)-(d) of Fig.~\ref{fig:gap_neumat}. These summarise the results for the gap at the Fermi surface in the singlet channel as a function of Fermi momentum. While we have access to the full momentum dependence of the gap, we restrict this discussion to the values at $k=k_F$ for brevity. The gap obtained in the BCS approximation with two-body-only forces and free single-particle spectra is shown in the panel (a). Solid, dashed and dotted lines correspond to three different N3LO chiral NN forces introduced in Refs.~\cite{Coraggio2013,Coraggio2014}. These are characterised by different cut-offs, ranging from $\Lambda=414$ MeV (solid lines) to $\Lambda=450$ MeV (dashed lines) and to the standard Entem-Machleidt force with $\Lambda=500$ MeV (dotted lines). Cut-off variation is one way of providing systematic errors, and the area covered by these 3 forces is therefore indicative of one type of uncertainty. We note that other errors, particularly those associated with different regulators for a given cut-off, should also be explored. 

\begin{figure}
\begin{center}
  \includegraphics[width=0.8\linewidth]{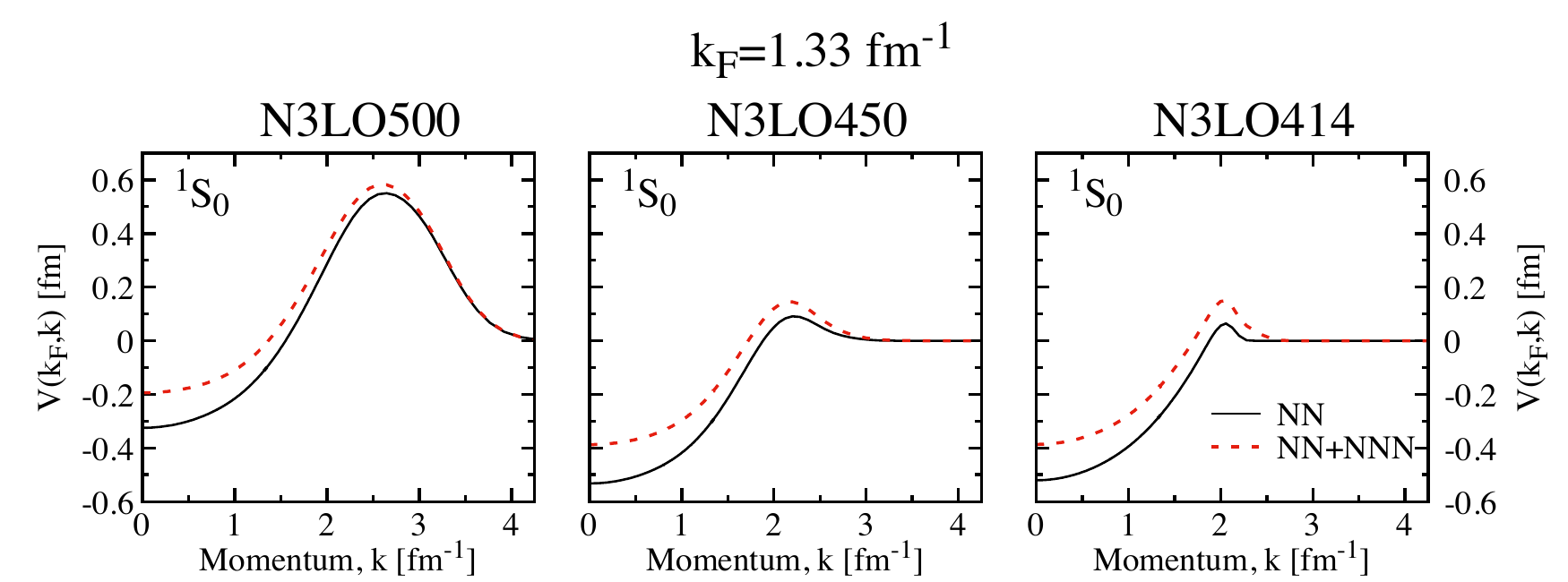}
 \end{center} 
\caption{Matrix elements of the pairing interaction $V(k_F,k)$ in the singlet channel in neutron matter at $k_F=1.33$ fm$^{-1}$. The solid lines in the left, central and right panels correspond to the N3LO500, N3LO450 and N3LO414 interactions. Dashed lines show the matrix elements when averaged 3NFs are introduced. }
\label{fig:NN_3N_1S0}       
\end{figure}

With these interactions, the maximum gap is located around $k_F \approx 0.86$ fm$^{-1}$ and ranges from $\Delta_\textrm{max} \approx 2.96$ MeV to $3.23$. In all cases, the gap closes somewhat above $k_F \approx 1.52$ fm$^{-1}$. This range of $\Delta_\textrm{max}$ is comparable to that associated with cut-off variations with semi-local chiral forces \cite{Drischler2017}. We note that the cut-off dependence around the maximum can be mostly ascribed to the N3LO450 interaction. The matrix elements of the pairing interaction at the Fermi surface, $V(k_F,k)$, for these three chiral forces are shown as solid lines in Fig.~\ref{fig:NN_3N_1S0}. We note that N3LO500 is significantly different in depth and shape from the N3LO450 and N3LO414 matrix elements. In particular, it extends to higher momenta as expected from a larger cut-off which is also implemented through a less sharp regulator \cite{Coraggio2014}. The differences between the N3LO450 and N3LO414 interactions are only visible near their corresponding cut-off momenta. These small differences should explain the observed variation of the gap in Fig.~\ref{fig:gap_neumat}.

The chiral results with NN forces agree well with those associated with traditional phase-shift equivalent interactions, like CD-Bonn \cite{Machleidt1995} and Argonne v18 (Av18) \cite{Wiringa1995}. We show these as dash-dotted and dash-double-dotted lines in Fig.~\ref{fig:gap_neumat}. We find that these two interactions provide results that are in between the N3LO414 and N3LO500 forces, peaking at $\Delta_\textrm{max} \approx 3.00$ MeV. One can therefore say that the gap in the singlet channel is relatively well constrained, at least within the lowest order BCS approach with NN forces only \cite{Dean2003,Drischler2017}.

In-medium effects should play an increasing role at higher densities. The first type of effect that we explore is that of 3NF. One of the advantages of chiral forces is that the associated 3NF can be derived on the same footing within the chiral perturbation theory expansion \cite{Epelbaum2009,Machleidt2011}. Here, we follow Ref.~\cite{Coraggio2014} and introduce 3NF at N2LO in the chiral expansion. The independent low-energy constants in the three-nucleon sector, $c_D$ and $c_E$, do not play a role in neutron matter. We note however that these have been adjusted to reproduce the experimental $^3$H and $^3$He binding energies and the Gamow-Teller component of the triton $\beta$-decay half-life \cite{Coraggio2014}. 

3NF matrix elements are predominantly repulsive in the relevant density regime for this channel. The matrix elements associated with the NN+3NF chiral forces are shown by dashed lines in Fig.~\ref{fig:NN_3N_1S0} at $k_F=1.33$ fm$^{-1}$. One clearly observes a repulsive shift due to 3NF throughout all momenta. It is therefore not surprising that the associated gaps in this channel become smaller when 3NF are included in a BCS calculation. Panel (b) in Fig.~\ref{fig:gap_neumat} shows that 3NF reduce the maximum gap to values in a range from $\Delta_\textrm{max} \approx 2.81$ to $\approx 2.96$ MeV. The effect of 3NF increases with density, and as a consequence, the closure of the gap is significantly affected by their presence. Whereas gaps based on NN forces only close around $k_F \approx 1.5-1.6$ fm$^{-1}$, gaps that include 3NF become zero below $\approx 1.4$ fm$^{-1}$. 

We now turn our attention to 
other in-medium effects. We have implemented two different approaches to treat this. Panel (c) shows results for an extension of the BCS approach in which the single-particle energies, $\varepsilon_k$, are obtained from a SCGF calculation. The pairing interaction is also quenched by a momentum-dependent $Z-$factor that is also obtained within SCGF. All results have been extrapolated to zero temperature. We find that the pairing gap is quenched by in-medium effects, and now the maximum gap lies in the region $\Delta_\textrm{max} \approx 2.40$ to $\approx 2.65$ MeV. The results from all NN interactions are very similar. Note, however, that this treatment does not take into account the full off shell dependence of the spectral strength.

A more realistic treatment of strength removal is obtained in a consistent implementation of Eq.~(\ref{eq:srcbcs}).
The open symbols in panel (d) of Fig.~\ref{fig:gap_neumat} correspond to the CDBonn and Av18 interaction SRC results already reported in Ref.~\cite{Ding2016}. SRC reduce the gap at maximum by about $0.5$ MeV with respect to the uncorrelated BCS approach and also influence the closure Fermi momentum, which is about $0.2$ fm$^{-1}$ smaller for SRC predictions compared to BCS. The inclusion of 3NF only changes these results quantitatively. Because the 3NF matrix elements are quite small around the maximum gap, their influence in the SRC-correlated results is mild around the maximum. At $k_F \approx 0.84$ fm$^{-1}$, the maximum gap with chiral NN+3NF ranges between $2.15$ and $2.34$ MeV. As a consequence, the results below about $k_F \approx 1$ fm$^{-1}$ are similar to those observed for CD-Bonn and Av18. Above this Fermi momentum, the gap with 3NF and SRC is somewhat smaller than the NN-only prediction. 

In general, we find results which are below the Av18 prediction by about $\approx 0.1-0.2$ MeV. We note that the origin of these small gaps is different in the case of Av18 and that of chiral forces. For Av18, SRC are significant at these densities as evidenced by the difference between the BCS and the SRC results. The same difference for the chiral NN+3NF forces is relatively smaller, which indicates that the effect of SRC is less relevant  for these soft chiral forces. Instead, small gaps are produced by the repulsive effect of 3NF. Finally, we stress that the gap closure when 3NF and SRC are considered occurs right above saturation, $k_F \approx 1.3$ fm$^{-1}$. Interestingly, the overall picture shows a relative agreement between all approaches, which suggests that the SRC-based depletion of the pairing gap is a rather universal effect. We also note that LRC are repulsive in the singlet channel and hence reduce the gap further \cite{Ding2016}.

\subsection{Triplet pairing}
\label{sec:ntriple}

The triplet channel in neutron matter has received attention recently \cite{Ding2016,Srinivas2016,Drischler2017,Dong2013} because of its potential relevance in neutron star cooling physics, particularly in the context of Cassiopeia A \cite{Page2011,Ho2015}. The bottom panels of 
Fig.~\ref{fig:gap_neumat} show the results of both BCS and SRC predictions for the neutron matter gap in this channel. Two striking features arise already at the BCS level in panel (e). First, unlike the singlet channel, there is a substantial deviation of predictions above Fermi momenta of order $k_F \approx 1.7$ fm$^{-1}$ in all the cases. At the lowest order BCS level, these differences  must be associated with the underlying pairing interaction. The attractive pocket of the diagonal partial waves is located at relatively large momenta, so much so that there is a very narrow range between the existing experimental data and pion creation threshold \cite{Srinivas2016}. In other words, phase shifts are not restrictive enough in this relatively high density region. Consequently, there is a wide range of predictions for pairing gaps based on NN-only calculations in the BCS approach [panel (e)]. Whereas CD-Bonn, Av18 and N3LO500 gaps peak above $0.5$ MeV somewhere above $k_F=2.2$ fm$^{-1}$, the softer chiral interactions N3LO450 and N3LO414 predict gaps with maxima below $0.3$ MeV at Fermi momenta below $2$ fm$^{-1}$. In fact, for the chiral interactions with a relatively low and sharp cut-off, the closure momenta are relatively close to the cut-off momenta.

\begin{figure}
\begin{center}
  \includegraphics[width=0.8\linewidth]{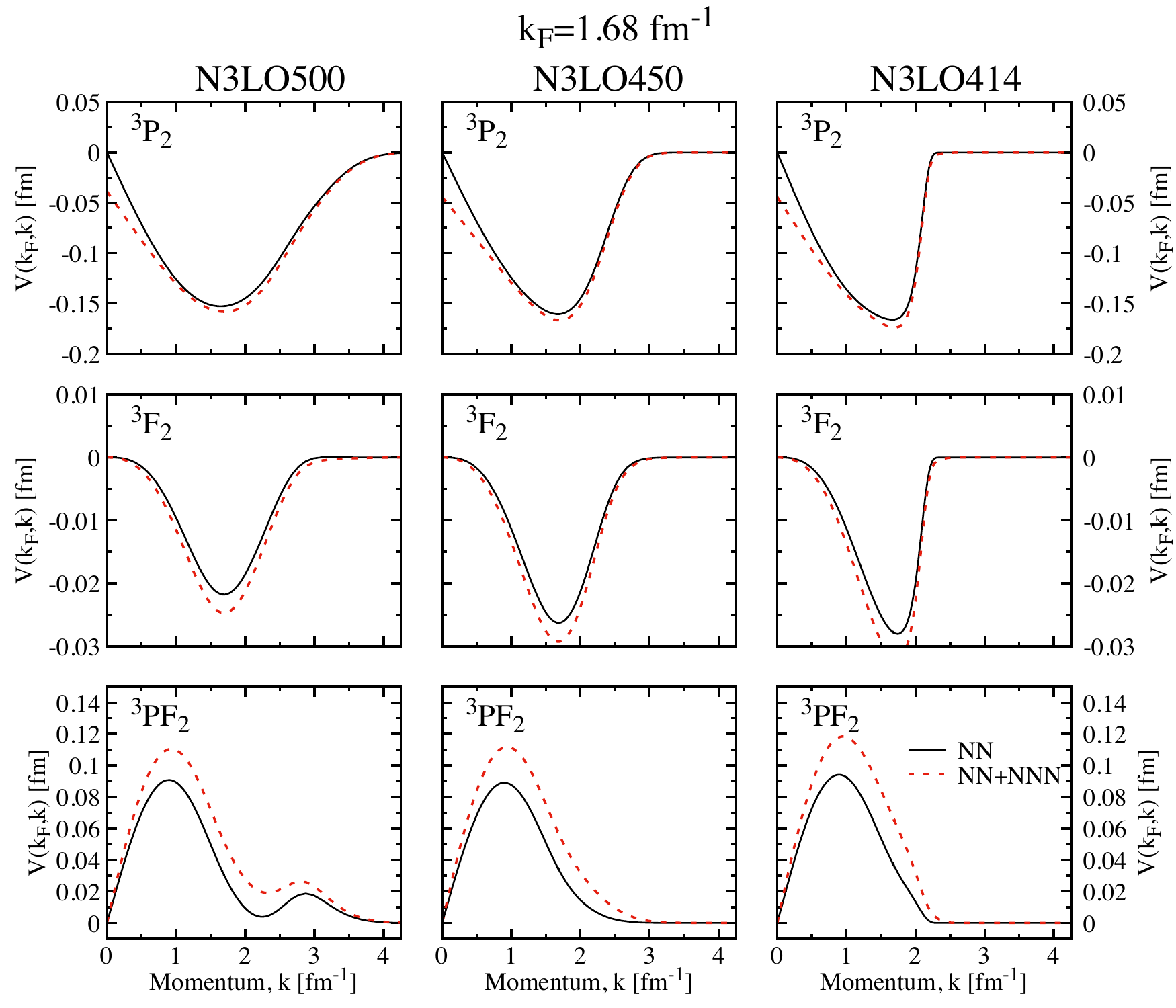}
 \end{center} 
\caption{The same as Fig.~\ref{fig:NN_3N_1S0} but at $k_F=1.68$ fm$^{-1}$. Top, central and bottom panels correspond to the partial wave channels $^3$P$_2$, $^3$F$_2$ and the coupling $^3$PF$_2$, respectively. }
\label{fig:NN_3N_3PF2}      
\end{figure}

The other relevant feature that we want to highlight at the BCS level is the effect of 3NF, shown in panel (f) of Fig.~\ref{fig:gap_neumat}. Whereas in the singlet channel 3NF played a repulsive role that reduced the gap, in the triplet channel 3NF actually tend to increase the gap. Again, at the BCS level this effect must be traced back to the pairing interaction. We plot the components of the pairing interaction $V(k,k_F)$ for the three relevant partial channels in Fig.~\ref{fig:NN_3N_3PF2} at $k_F=1.68$ fm$^{-1}$. The effect of 3NF in the pairing interaction in the diagonal $^3$P$_2$ and $^3$F$_2$ channels is attractive, but relatively mild. 3NF are larger in the coupled channels, where they are repulsive in absolute value. Overall, the attractive 3NF components must be dominant and result in a larger gap at all densities for each force. With 3NF, the gaps reach maximum values of $\Delta_\textrm{max}=0.5-0.6$ MeV for N3LO414 and N3LO450, and $1.22$ MeV for N3LO500. The gap closure for the two lower cut-off forces is not affected much by 3NF, which suggests that the closure is due to the presence of the cut-off independently of 3NF. 

The effect of SRC is significantly different in the triplet channel than in the singlet channel. 
Again, we have implemented two different approaches for SRC - an incomplete treatment using single-particle energies and renormalization factors (or ``BCS+$\varepsilon_k$+$Z_k$" approximation) [panel (g)] and a complete treatment of fragmentation in pairing physics effects [panel (h)]. 
In either approach, CD-Bonn and Av18 gaps are strongly suppressed at high densities, and the maximum gaps are well below $0.1$ MeV in the more complete treatment of SRC. These relatively hard forces have significant high-momentum components that remove effectively strength around the Fermi surface. With an interaction that is only marginally attractive, this gives rise to a large damping of the pairing gap. We note that the slightly erratic nature of the calculations in panel (h) is due to the zero-temperature extrapolation procedure of the denominator in Eq.~(\ref{eq:fragden_normal}), which is particularly sensitive to very narrow structures around the Fermi surface. The BCS+$\varepsilon_k$+$Z_k$ approximation is also obtained from a zero-temperature extrapolation, but no convolution integrals are involved. The results are thus numerically more stable.

Chiral NN+3NF calculations including SRC between $k_F=1.5$ and $1.8$ fm$^{-1}$ predict gaps that increase with density and lie in the range between $0.1$ ($0.1$) and $0.3$ ($0.5$) MeV in the SRC (BCS+$\varepsilon_k$+$Z_k$) treatment. These are significantly larger than the CD-Bonn and Av18 results due to the combined effect of attractive 3NF and an overall smaller effect of SRC for these renormalized, soft interactions. Above $k_F \approx 1.85$ fm$^{-1}$, the predictions based on chiral forces diverge substantially. The closure of the N3LO414 gap is abrupt and in the region $k_F \approx 1.85-2.05$ fm$^{-1}$, below its corresponding BCS value. N3LO450 yields the largest maximum gaps when SRC are considered fully, up to about $\Delta_\textrm{max}=0.45$ MeV. We note that this is close to the BCS value, which indicates that the effect of the fragmented energy denominator is not necessarily associated with a damping of the gap, although this happens in a regime where the applicability of this soft chiral force is questionable. Finally, the results based on N3LO500 with 3NF have not reached a maximum gap up to about $k_F=2.2$ fm$^{-1}$. This is the maximum density at which our extrapolation procedure works reliably and therefore there is no observable closure in panels (g) and (h) of Fig.~\ref{fig:gap_neumat}.

\section{Pairing in symmetric nuclear matter}
\label{sec:symmetric}

\begin{figure}
  \includegraphics[width=\linewidth]{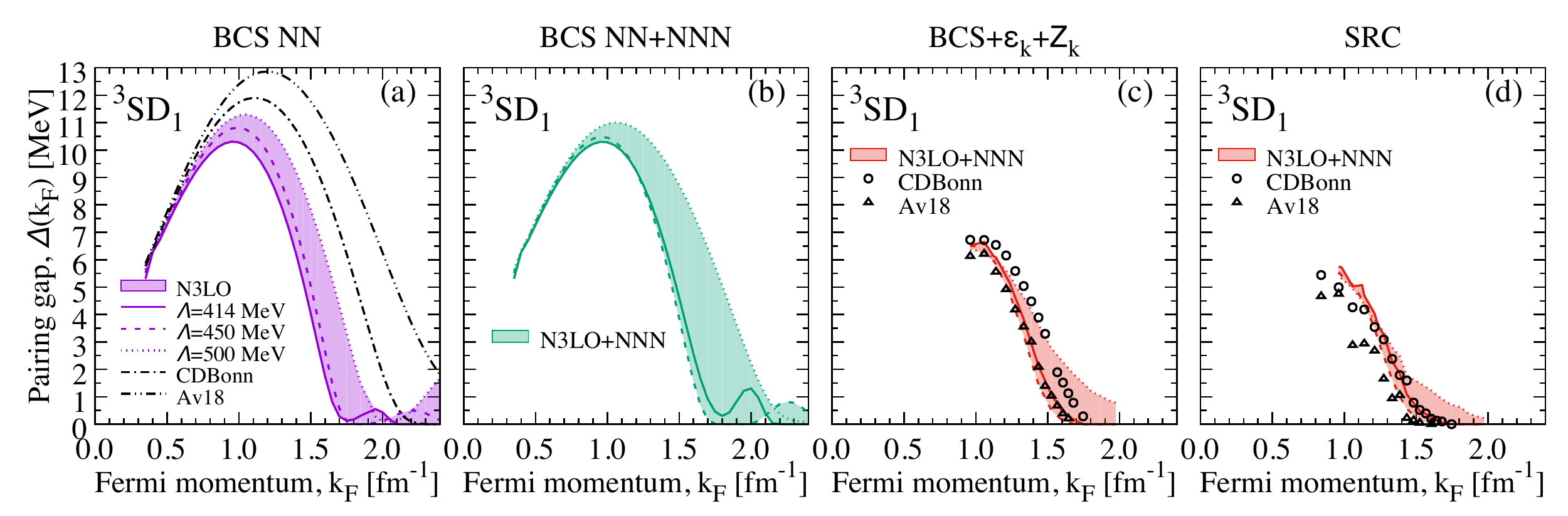}
\caption{Triplet gap in nuclear matter at the Fermi surface as a function of Fermi momentum, $k_F$. The different panels correspond to (a) BCS results with NN forces only,  (b) BCS results with chiral NN+3NF forces, (c) BCS+$\varepsilon_k$+$Z_k$ results with both NN and chiral NN+3NF forces and (d) SRC results with NN forces and chiral NN+3NF forces, respectively.} 
\label{fig:3SD1}      
\end{figure}

We now turn our attention to the triplet gap in symmetric nuclear matter. This is operated by the attractive $^3$SD$_1$ partial wave which is responsible for the existence of the deuteron and also, to a large extent, for the saturation of nuclear matter \cite{Dewulf2003}. 
Our results are summarised in the four panels of Fig.~\ref{fig:3SD1}. 
Panel (a) focuses on BCS results with NN-only interactions, including chiral forces as well as CDBonn and Av18. The BCS results predict very large gaps in all NN-only calculations in agreement with previous studies \cite{Vonderfecht1991,Baldo1992,Baldo1995,Maurizio2014}. Av18, for instance, has a maximum gap of $\Delta_\textrm{max} \approx 13$ MeV around a Fermi momentum, $k\approx 1.3$ fm$^{-1}$, which is very close to saturation density. The BCS gap for CDBonn is only $1$ MeV smaller, but its closure occurs significantly earlier at $k_F \approx 2.2$ fm$^{-1}$. 

BCS calculations with chiral NN forces confirm the qualitative picture of large maximum gaps, in the region $\Delta_\textrm{max} \approx  10-11.5$ MeV.  These maximum gaps occur at slightly lower densities than the CDBonn and Av18 counterparts, $k_F \approx 1$ fm$^{-1}$. Beyond this density, the gaps steadily decrease to a minimum at around $k_F = 2$ fm$^{-1}$. Surprisingly, the gaps do not close here and, in the region beyond $k_F > 2$ fm$^{-1}$, we find gaps that increase with density. However, this happens in an area where the matrix elements are likely to be sensitive to regulator effects and hence we do not necessarily expect these results to reflect any real physics.

\begin{figure}
\begin{center}
  \includegraphics[width=0.8\linewidth]{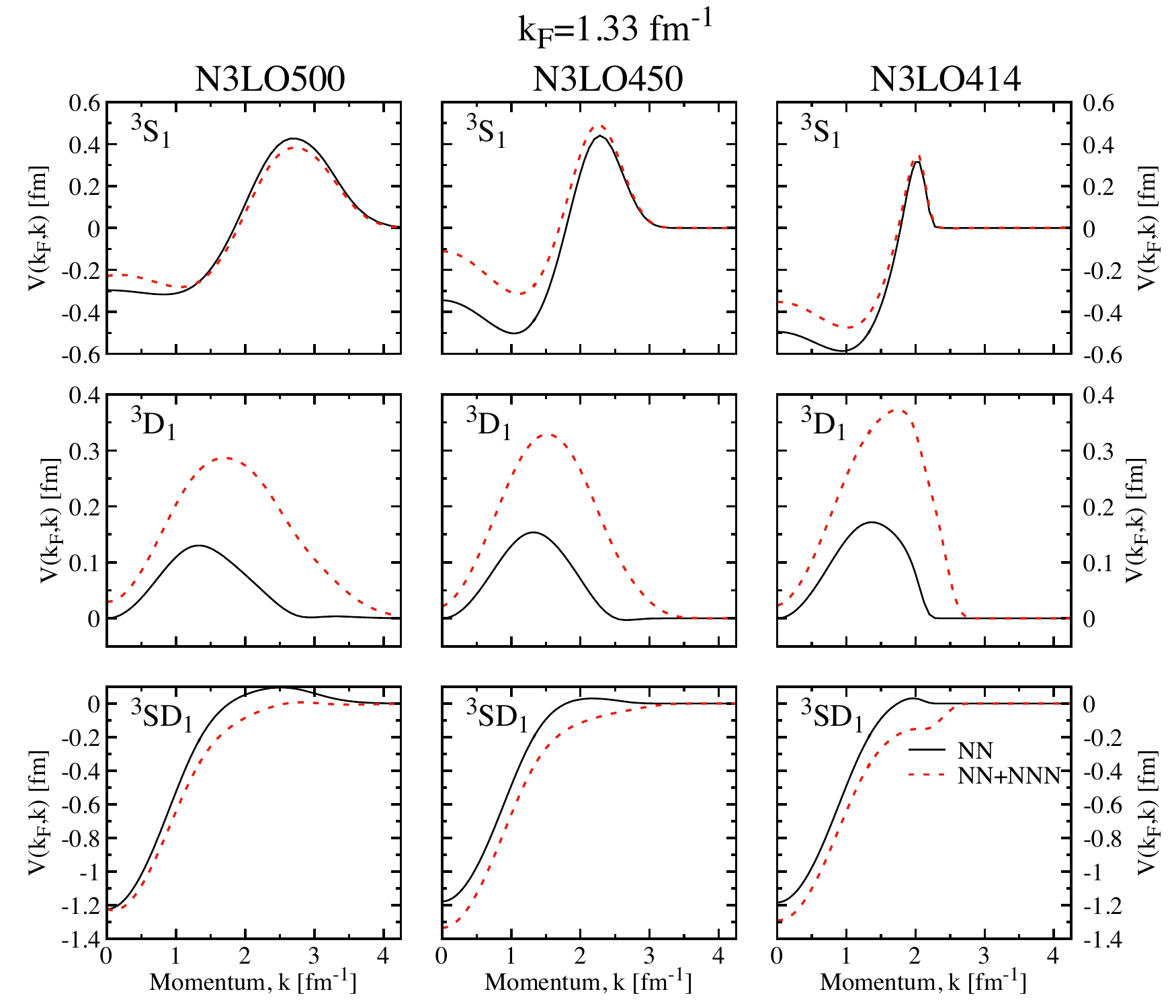}
 \end{center} 
\caption{The same as Fig.~\ref{fig:NN_3N_1S0} but at $k_F=1.33$ fm$^{-1}$ for symmetric nuclear matter. Top, central and bottom panels correspond to the partial wave channels $^3$S$_1$, $^3$D$_1$ and the coupling $^3$SD$_1$, respectively. }
\label{fig:NN_3N_3SD1}
\end{figure}

Figure 4.b shows BCS results with NN and 3NF forces. 3NF in the triplet channel of nuclear matter can provide an effective saturation mechanism of nuclear matter \cite{Carbone2013b,Coraggio2014}. Because they significantly alter the nuclear interaction in symmetric matter, one could naively expect relatively large effects in the corresponding pairing gaps. In contrast, the associated BCS pairing gaps including 3NF essentially overlap the results obtained with NN forces up to about the saturation Fermi momentum $k_F \approx 1.3$ fm$^{-1}$. The pairing matrix elements at $k_F \approx 1.3$ fm$^{-1}$ with (dashed lines) and without 3NF (solid lines) are shown in Fig.~\ref{fig:NN_3N_3SD1}. 3NF are mildly repulsive in the diagonal $^3$S$_1$ channel for N3LO450 and N3LO414, and they are very small for N3LO500. In contrast, the 3NF are clearly repulsive in the diagonal D wave channel. The coupling in the off-diagonal channel is unambiguously attractive. All in all, the contributions at saturation density are relatively small and one cannot disentangle easily whether they have a repulsive or attractive nature. 

The situation appears to change at suprasaturation densities. There, the gaps for the N3LO500+3NF interaction are a few MeV larger than the associated NN-only force in the region $k_F=1.5-2$ fm$^{-1}$. This is in contrast to the N3LO414 and N3LO450 cases, for which the 3NF contributions to the gap are attractive, but relatively small (between $0.5-2$ MeV). We also find that 3NF tend to reduce the appearance of second nonzero gap area at large densities. 

The effects of SRC in the gap are very significant. 
Panel (c) shows the results for the BCS+$\varepsilon_k$+$Z_k$ approximation. All results indicate a maximum gap of around $\Delta_\textrm{max} \approx 6.2-6.7$ MeV at $k_F \approx 0.95-1.05$ fm$^{-1}$. The closure for CD-Bonn and Av18 occurs between $1.7$ and $1.8$ fm$^{-1}$, in relative agreement with N3LO414+3NF and N3LO450+3NF results. The closure for the N3LO500+3NF gap occurs beyond $2$ fm$^{-1}$.

The results for panel (d) correspond to a more consistent implementation of SRC. 
For both the traditional phase-shift equivalent forces and the chiral interactions, it appears that the maximum of the gap is reduced to about $\Delta_\textrm{max} \approx 4.7-5.5$ MeV.
In other words, SRC reduce the maximum gap to about $50 \, \%$ of the BCS maximum value. These results are in agreement with the preliminary calculations using this very same method but somewhat different numerical techniques in Ref.~\cite{Muther2005}. Moreover, the SRC gap associated with chiral NN+3NF depends mildly on the cut-off value and the implementation of the SRC effect, which suggests that the strong gap reduction is a relatively generic feature that is insensitive to many-body uncertainties. 

We also find that the gap closure with SRC occurs at significantly lower densities than the corresponding BCS results. With CDBonn and Av18, the closure occurs around  $k_F \approx 1.5-1.7$ fm$^{-1}$. N3LO414+3NF and N3LO450+3NF provide very similar closures, but N3LO500+3NF appears to level off at high densities and a crude extrapolation indicates a closure closer to $k_F \approx 2$ fm$^{-1}$. This corresponds to densities above $1.5$ times saturation. Our calculations suggest that SRC do deplete neutron-proton pairing correlations significantly, but not enough to completely remove them in this channel. We note that our results with chiral interactions include 3NF consistently with the underlying many-body approach, and therefore a different treatment of the 3NF will not  change this picture significantly. The situation regarding LRC is difficult to address because of the interplay of spin and isospin excitations. Preliminary calculations in Ref. \cite{Cao2006} indicate that LRC might provide an antiscreening effect that could enhance the gap, but the effect of the strong tensor channel is difficult to quantify in this context \cite{Dickhoff1981,Dickhoff1981x}. 

\section{Conclusions and future prospects}
\label{sec:conclusions}

Superfluid nuclear systems require special many-body treatments due to the interplay between a strongly repulsive core, the presence of attractive components at intermediate distance and a strong tensor coupling \cite{Dean2003}. The complex structure of the NN force induces many-body correlations that have to be accounted for in the treatment of pairing \cite{Dickhoff08}. It is possible that these correlations cannot be accounted for realistically within a BCS picture and that more consistent and realistic approaches are needed to describe superfluidity in nuclear physics. 

Here, we have presented results obtained in a Gorkov-inspired treatment that consistently takes into account the SRC arising from the strong NN force. We have presented results for neutron matter pairing in the singlet and triplet channels, as well as for the tensor channel in symmetric nuclear matter. All our results indicate that SRC deplete the gap from its BCS value. In the singlet neutron matter channel and in the triplet symmetric matter channel, we find that results from traditional phase-shift equivalent NN forces as well from chiral NN+3NF calculations are in relatively good agreement. In neutron matter, the maximum gap decreases by about $0.5$ MeV in all cases after SRC are introduced. In symmetric matter, the maximum gap at the BCS level is large, above $10$ MeV. SRC, in contrast, reduce the maximum gap by about $50 \, \%$ independently of the NN force and of the specifics of the 3NF that are considered. SRC also tend to reduce the gap closure density, although the details are sensitive to the structure of the NN force. 

The triplet neutron matter channel is much more sensitive to the underlying NN force. At the BCS level, the large differences obtained in maximum gaps and closures are related to the underlying NN force, which is not anymore constrained by phase-shift analysis at the densities where the triplet channel is active. Further, there are significant differences in the density dependence of SRC at large neutron matter densities. As a consequence, the SRC gaps are relatively different depending on the force under consideration. While SRC results with traditional NN forces yield gaps well below $0.1$ MeV in the channel, chiral forces can result in maximum gaps of the order of $0.5$ MeV even after SRC are considered. Having said that, chiral forces are at the limit of their applicability in this large density regime, as evidenced by the large cut-off dependence. 

Future advances in the treatment of correlations in nuclear pairing will go along two directions. A more consistent treatment of SRC can be achieved by implementing fully self-consistent Gorkov-Green's functions calculations that do not rely on zero-temperature extrapolations \cite{Bozek1999}. The numerical tools exist and have been implemented in nuclear structure already \cite{Soma2014}. 
In infinite matter, the calculation of the superfluid spectral function in a fully consistent theory can be difficult. Similarly to the normal case, the superfluid spectral function has two very different features: narrow peaks close to the Fermi surface and tails that extend to high momenta and energies. These preclude the application of standard techniques with uniformly spaced grids in energy and momentum. These structures will in turn affect the normal phase via a feedback mechanism, so a fully self-consistent calculation will necessarily involve a recalculation of the normal properties at each density. We are not aware of any successful attempts to implement the Gorkov formalism in infinite matter using fully realistic NN and 3N forces. This would however provide invaluable information on the interplay between both SRC and LRC in pairing properties via a perturbative expansion. 
We note however that the effect of pairing is still mostly restricted to momenta near the Fermi momentum and is expected to have little effect on the normal self-energy as for example has always been argued by Migdal ~\cite{Migdal1967}.

Another relevant aspect which we have ignored her for brevity is the contribution of LRC to the effective pairing interaction. Depending on the system, the density and the many-body approach, the pairing interaction can be screened or antiscreened by the exchange of spin, density and isospin fluctuations in the medium. A consistent treatment in which these correlations are obtained from an underlying NN force is still missing and should be a first priority to elucidate extremely relevant questions associated with nuclear superfluids. 

\begin{acknowledgements}
This material is based upon work supported by 
STFC through Grants ST/I005528/1, ST/L005743/1  and  ST/N002636/1;
Grant No. FIS2014-54672-P from MICINN (Spain); Grant No. 2014SGR-401 from Generalitat de Catalunya (Spain);
and NSF Grant PHY-1613362. 
Partial support comes from ``NewCompStar" COST Action MP1304. 

\end{acknowledgements}

\bibliographystyle{spphys}       
\bibliography{biblio}   

\end{document}